# Plasmonic structure integrated superconducting nanowire single-photon detector constructed with BSCCO patterns

András Szenes[1,3], László Pothorcki[1], Balázs Bánhelyi[2,3], and Mária Csete[1,3,*]

*1. Department of Optics and Quantum Electronics, University of Szeged, 6720, Dóm tér 9, Szeged, Hungary.*
*2. Department of Computational Optimization, University of Szeged, 6720, Árpád tér 2, Szeged, Hungary*
*3. HUN-REN Wigner Research Center of Physics, Konkoly-Thege Miklós út 29-33, Budapest, Hungary.*
**mcsete@physx.u-szeged.hu*

**Keywords**: single-photon detector, numerical optimization, plasmonics, nanocavity, cuprate, effective medium

**Abstract:** Superconducting nanowire single-photon detectors (SNSPDs) were designed by integrating periodic plasmonic structures onto hBN protected superconducting BSCCO patterns to enhance the absorptance. The numerical investigation of optimized nanocavity-array (NCAI) and nanocavity-trench-array (NCTAI) SNSPDs has revealed that more than one order of magnitude larger absorptance can be achieved at perpendicular incidence, when compared to the corresponding meandered BSCCO pattern in a resonant nanocavity. The presented SNSPDs are considerably improved via first and third quarter wavelength nanocavity resonances, as evidenced by the near-field maps and validated by the standard retrieval method. Although, NCAI-SNSPD exhibits a slightly larger absorptance, NCTAI-SNSPD remains competitive due to its smaller filling factor, thereby allowing for smaller charge crowding effect.

**Introduction:** The detection of single-photons encompasses numerous fields of fundamental research and applications, notably in the secure information transfer and has a crucial role in quantum computing and quantum communication [1-3]. Among state-of-art single-photon detectors, superconducting nanowire single-photon detectors (SNSPDs) have remarkable properties [4]. These detectors operate based on the modification of the superconducting state in the nanowires constituting their absorbing segments, as the single-photon induced hot spot is capable of breaking the superconducting equilibrium. The consequent increase beyond the critical current manifests itself in a voltage jump in the readout electronics [5]. Exploiting this phenomenon, SNSPDs offer operation across a wide range of wavelengths (from X-ray to mid-IR) and enable high efficiency as well as picosecond time-resolution [6-9]. A drawback of SNSPDs is that their operation is restricted to very low temperatures, typically below 4.2 K, in the case of e.g. niobium-nitride (NbN) superconductors. This cryogenic temperature requirement constrains their wide-range applicability and made them very expensive.

Efforts have been made to develop high-critical-temperature cuprate superconductors for SNSPDs, yet experimentally detecting single-photons above 4.2 K was not successful [10, 11]. The primary challenge is their manufacturing, as the traditional SNSPD fabrication procedures severely damage cuprate superconductors. Recently, a method was developed, whereby non-superconducting patterns are drawn into helium-ion-exposed $Bi_2Sr_2CaCu_2O_{8+\delta}$ (BSCCO) flakes encapsulated into hBN, thus achieving desired patterns without damaging the superconducting material. Using this fabrication method, thin BSCCO flakes were decorated with complementary superconducting wires that reproducibly demonstrated single-photon sensitivity at telecommunication wavelengths (1550 nm) under operation temperatures of 20–25 K.

These detectors achieved a detection efficiency (DE), that is governed by the in-coupling efficiency, absorption and voltage generation probability, in the order of DE~$10^{-4}$. Their dynamical behavior is characterized by nanosecond-scale recovery times, while their dark count rate (a reference electronic response without detectable optical signal) is approximately $10^3$ counts per second, proving their potential for practical applications in single-photon detection [12, 13]. Thus-formed novel generation SNSPDs can be operable with simpler cooling systems; however, the periodic pattern size is currently limited to a few tens of micrometers due to the degradation of BSCCO during the fabrication process [12, 13].

The next challenge is enhancing the absorption efficiency of the nanowire pattern. One of the previously demonstrated approaches is embedding the superconducting nanowires into an optical cavity [14, 15]. Our previous studies have shown that in addition to continuous optical nanocavities, significant improvements can be achieved by integrating periodic nanoplasmonic patterns into NbN based SNSPDs [16-19]. Integration of plasmonic patterns enables the application of relatively high (wavelength-scaled) periodic nanowire patterns in a wide wavelength-range and within a specific range of incidence angles, while maintaining high efficiency (>90%) [20-22]. The large period allows for reduced kinetic inductance, if the pattern size remains fixed, as the total nanowire length decreases. This effect is particularly relevant for NbN-based patterns, where the high kinetic inductance has a significant impact on the detector speed [23], though it is not crucial for BSCCO patterns of two orders of magnitude lower kinetic inductance. By integrating l various previously demonstrated methods, nanocavities could be fabricated using conventional lithography, while the hBN protected BSCCO pattern, which could be processed with He-ion lithography, could be transferred by ensuring synchronization with the nanocavity-array [12-13].

Accordingly, in this study, the numerical optimization of BSCCO-based SNSPDs integrated with plasmonic patterns was performed to maximize the absorptance in the nanowires, thereby determining the optimal geometry for efficient operation at telecommunication wavelengths in the C-band. The higher operational temperature and high efficiency make these SNSPDs suitable for integrated photonic circuits, moreover as on-chip optical components they pave the way for novel quantum optical applications.

**Methods:** The geometries of SNSPDs were designed in a numerical environment, with the RF module of FEM-based COMSOL Multiphysics software package, by solving the full-wave Maxwell equations, considering the material's dispersion by importing tabulated datasets [24-25]. The optical response of different device compositions, specifically the absorptance (arising as a loss from resistive heating in materials with non-zero imaginary parts of their permittivity), reflectance, and transmittance (computed as the outgoing power density integrated on port-boundaries in the forward and backward directions), were determined as a function of wavelength and angle of incidence, by using single unit cells of the inspected periodic patterns, terminated vertically by Floquet boundary conditions.
The superconducting material segment was a 15 nm thick BSCCO layer (calculations supposing 30 nm thick wires are presented in the Supplementary Materials). Superconducting BSCCO wires were separated by interleaving residual BSCCO, arising after etching and acting as an insulator (reBSCCO). During modelling, this reBSCCO was considered so as the imaginary part of permittivity of the insulator BSCCO layer was set to near zero. The complete BSCCO layer was covered by a 20 nm hBN film, and both were defined by their full permittivity tensors accounting for the thin films anisotropy [26-27] (Figure S1 in Supplementary Materials).

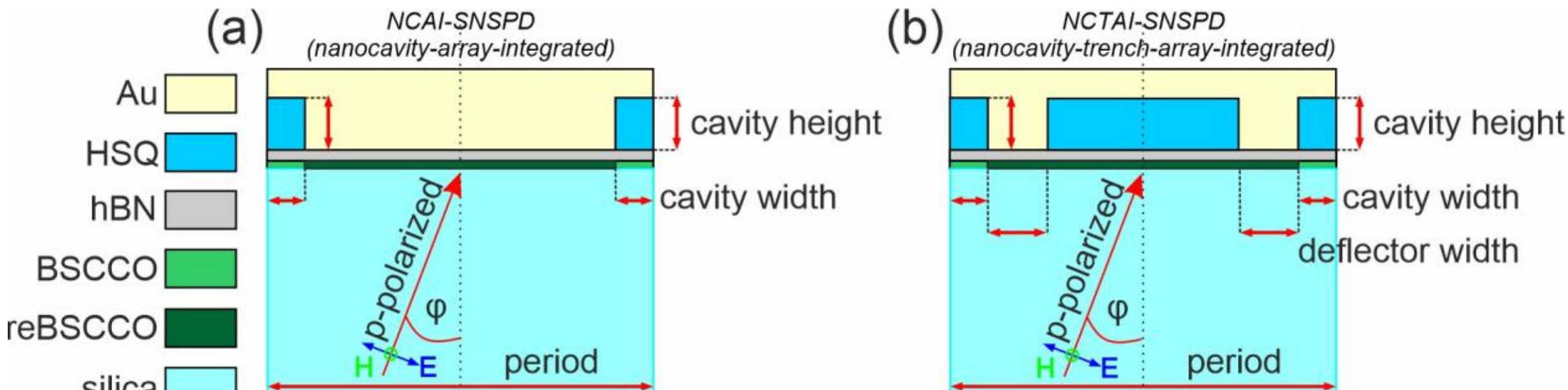

***Figure. 1.*** *Schematic of the unit cells of the inspected NCAI-SNSPD (left) and NCTAI-SNSPD (right) periodic structures, indicating the illumination by p-polarized light and the parameters varied during optimization.*

The optimization of the geometry was realized to maximize the amount of power absorbed in the BSCCO wires (BSCCO absorptance) at 1550 nm, which is important wavelength for telecommunication applications. Normal incidence was supposed during the optimization performed by using p-polarized light in S-configuration [20].

During this study two types of integrated SNSPD device compositions were inspected (Figure 1). In both compositions the 15 nm thick meandered BSCCO wire is positioned onto a silica substrate and covered by the 20 nm hBN and a Hydrogen silsesquioxane (HSQ) layer, with a thickness varied in the [10 nm, 800 nm] interval during optimization. On the top of the HSQ layer a fixed 60 nm thick gold reflector closes an optical nanocavity that is filled with HSQ. In the first inspected device composition the superconducting wires were separated by bulk gold segments laterally, forming a nanocavity-array (NCAI – nanocavity-array-integrated SNSPD), with a wire-to-wire distance that was varied in the [800 nm,1200 nm] interval. In the second device composition the width of the lateral walls of the cavities was also varied symmetrically, thereby allowing for "opening an empty trench array". The HSQ filled cavities and trenches create a nanocavity-trench-array (NCTAI – nanocavity-trench-array-integrated) SNSPD. Both inspected composition types are inherited from the integrated devices efficiently optimized in our previous studies to improve NbN-based SNSPD absorptance, when it was shown that by introducing a secondary nanocavity (trench) array, even greater absorptance can be achieved, due to the minimized competitive gold absorptance [20-22]. The geometrical parameters of both compositions were optimized using COMSOL Multiphysics Optimization Module and an in-house developed GLOBAL algorithm [28].

During optimization as a first step, a Monte Carlo simulation was conducted to determine the period, nanocavity height, (deflector width) and wire width in NCAI (NCTAI) SNSPD, with a potential to maximize the achievable absorptance. As a second step a local search was performed to find the global maximum in the inspected geometrical parameter region. The objective function was defined to maximize BSCCO absorptance. The far-field and near-field responses of the optimized device, as well as the optical response including the simulated absorptance that correlates with the detection efficiency across the inspected spectral range, were thoroughly investigated for the BSCCO devices to uncover the underlying physics. The absorptance of optimized devices was compared to those of standalone BSCCO meander in simple optical cavity (Figure S2-S3 and Table S1 in the Supplementary Materials),

Thereafter the optimized devices were inspected in the 1000-2000 nm band applying tilting as well, to map their dispersion characteristic. The dispersion characteristics of the two integrated device compositions taken in p-to-S configuration were compared to each other in order to characterize their potential as efficient and fast single-photon counting devices at 1550 nm.

## Results and discussion

**1. Reference configurations:** To define a reference for our optimized nanostructure-integrated SNSPDs comparable to the optimized geometries, compositions involving (i) a periodic, 100 nm wide BSCCO meander on silica substrate, (ii) covered with an optical cavity were first investigated (Figure S2 and S3 in the Supplementary Materials).

In the case of a periodic BSCCO meander, small 0.014 BSCCO absorptance is achieved at 1550 nm at perpendicular incidence, which is accompanied by large 0.956 transmittance (Figure S2). The absorptance can be slightly increased by increasing the operation wavelength, due to the increasing imaginary part of BSCCO permittivity. Oblique incidence is also advantageous in this configuration, as it is demonstrated by reaching 0.019 absorptance at 58° angle of incidence, which is promoted by the propagating mode on the BSCCO slab.

The optical cavity (OC) is formed by placing a 60 nm gold reflector above the BSCCO meander (Figure S3). To maximize absorptance, the cavity length is tuned to 225 nm. At 1550 nm and at perpendicular incidence, 0.041 absorptance is achieved, which is accompanied by nearly zero transmittance, a minimal 0.027 competitive Au absorptance, and a high, 0.921 reflectance. The absorptance can be slightly increased by increasing the operation wavelength; however, tilting is not advantageous in this composition.

**2. Nanocavity array:** Dividing the optical cavity with vertical gold segments creates a nanocavity array, which is the common element of the proposed plasmonic structure integrated SNSPDs (Figure 1).

According to the initial Monte Carlo simulations in both the nanocavity-array-integrated (NCAI-SNSPD) and the nanocavity-trench-array-integrated (NCTAI-SNSPD) compositions, two absorptance peaks emerge as a function of nanocavity length, that correspond to the $\lambda/(4\times n_{eff})$ and $3\times\lambda/(4\times n_{eff})$ modes of the nanocavities (Figure S4a and S5a).

In the case of NCAI-SNSPD, absorptance plateaus were observed for larger cavity widths and smaller periods i.e., corresponding to increasing filling factor (Figures S4b and c). The result of plateauing indicates that a significant reduction in period or increase in wire width is not accompanied by a significant increase in absorptance. A trade-off exists for this composition, therefore using an intermediate period, allowing for a reduction in charge crowding effect, with only a slightly compromised decrease in absorptance, is reasonable to consider.

In the case of NCTAI-SNSPD, a global maximum emerges at small /large /intermediate cavity width/period/deflector width, respectively (Figure S5 in the Supplementary Materials).

Following Monte Carlo simulations, local searches were realized with NCAI-SNSPD and NCTAI-SNSPD around the two absorptance peaks corresponding to different nanocavity resonant modes (Figures 2 and 3).

The NCAI-SNSPD optimized to resonate in the $\lambda/(4\times n_{eff})$ cavity mode exhibits a broad Lorentzian-resonance in absorptance centered around 1550 nm (Figure 2a solid lines). The peak absorptance of NCAI-SNSPD is 0.726, which is accompanied by a small 0.039 reflectance and negligible transmittance, and 0.216 competitive gold absorptance. The BSCCO absorptance reaches a global maximum at perpendicular incidence, as indicated by the polar angle-dependent optical response, at the center of a pass band (Figure 2b solid lines). In the close vicinity of the BSCCO nanowire 8.3-fold maximum near-field enhancement is reached (Figure 2c, left).

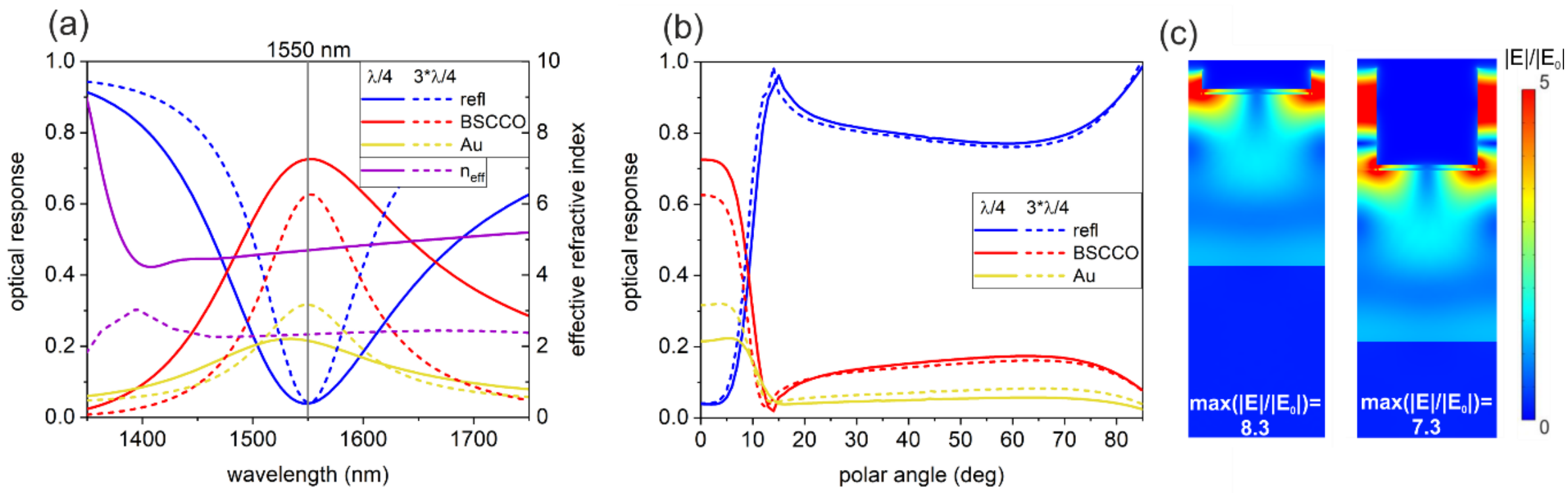

***Figure 2**. Optical response of optimized NCAI-SNSPD composition in p-to-S configuration. (a) optical response and effective refractive index spectrum and (b) polar angle dependence. Solid: ~$\lambda/(4\times n_{eff})$ mode dashed: ~$3\times\lambda/(4\times n_{eff})$ mode. (c) time-averaged near-field enhancement of NCAI-SNSPD optimized to resonate in the (left) $\lambda/(4\times n_{eff})$ mode and (right) $3\times\lambda/(4\times n_{eff})$ mode. Please note that the color scale was saturated at a value of 5 to enhance the visibility of fine structural details; the actual maximum near-field intensity exceeds this value.*

When optimized for the cavity mode resonating in a nanocavity of ~$3\times\lambda/(4\times n_{eff})$ length, the NCAI-SNSPD shows a smaller and narrower absorptance peak centered around 1550 nm (Figure 2a dashed lines). The higher Q-factor is typical for higher-order plasmonic resonances. The maximal absorptance is 0.626, accompanied by a small, but non-negligible reflectance (0.041) and a higher gold absorptance (0.317), when compared to those achieved in NCAI-SNSPD optimized for resonance in the $\lambda/(4\times n_{eff})$ cavity mode. The polar angle dependence of the optical responses in case of the higher-order mode closely resembles that monitored in case of the $\lambda/(4\times n_{eff})$ cavity resonant mode, with consistently smaller absorptance values and extrema occurring at slightly smaller polar angles (Figure 2b dashed lines). The highest absorptance is achieved at perpendicular incidence, similarly at the center of a pass-band, which is slightly narrower. A 7.3-fold maximum near-field enhancement is achievable, which is slightly smaller compared to that reached in NCAI-SNSPD optimized to resonate in the $\lambda/(4\times n_{eff})$ cavity mode (Figure 2c).

Standard retrieval calculations [29] indicate that the optimal cavity lengths correspond to approximately $\lambda/(4\times n_{eff})$ and $3\times\lambda/(4\times n_{eff})$ with good precision (Figure 2a, green lines).

The NCTAI-SNSPD is optimized to resonate in the $\lambda/(4\times n_{eff})$ and $3\times\lambda/(4\times n_{eff})$ modes as well (Figure 3). In case of the lower order mode the absorptance spectrum of the optimized NCTAI-SNSPD reveals a similar pass-band centered at 1550 nm as it was shown in the optimized NCAI-SNSPDs (Figure 3a solid lines).

The global maximum in the absorptance spectrum is 0.715, which is accompanied by a small reflectance (0.059), negligible transmittance, and a relatively large competitive gold absorptance (0.214). This maximum is attained at perpendicular incidence (Figure 3b solid lines). At larger polar angles, the absorptance rapidly decreases, while at intermediate values, it exhibits a wider, shallower local absorptance maximum. At the global maximum an 8-fold near-field enhancement is observable (Figure 3c).

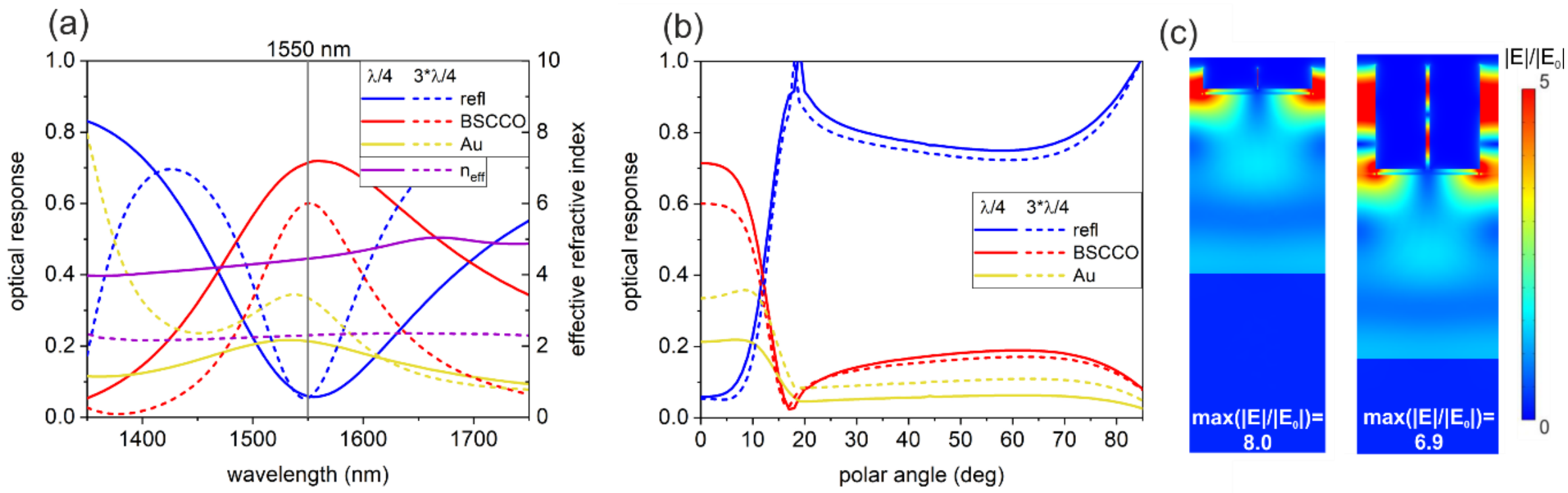

***Figure 3****. Optical response of optimized NCTAI-SNSPD composition in p-to-S configuration. (a) optical response and effective refractive index spectrum and (b) polar angle dependence. Solid: ~λ/(4×n*$_{eff}$*) mode dashed: ~3×λ/(4×n*$_{eff}$*) mode. (d) time-averaged near-field enhancement of NCTAI-SNSPD optimized to resonate in the (left) λ/(4×n*$_{eff}$*) mode and (right) 3×λ/(4×n*$_{eff}$*) mode. Please note that the color scale was saturated at a value of 5 to enhance the visibility of fine structural details; the actual maximum near-field intensity exceeds this value.*

When optimized to resonate in the $3\times\lambda/(4\times n_{eff})$ mode, the NCTAI-SNSPD exhibits a significantly narrower peak in the absorptance spectrum, which is precisely centered at 1550 nm (Figure 3a dashed lines). The maximal absorptance is 0.601, accompanied by a small (0.054) reflectance and a more significant competitive gold absorptance (0.335). This maximum is reached at perpendicular incidence, with the absorptance rapidly decreasing at small tilting, while a small local maximum appears at large polar angles (Figure 3b dashed lines). The near-field enhancement at the global maximum is 6.9-fold, which is slightly smaller compared to that reached in NCTAI-SNSPD optimized to resonate in the $\lambda/(4\times n_{eff})$ cavity mode (Figure 3c).

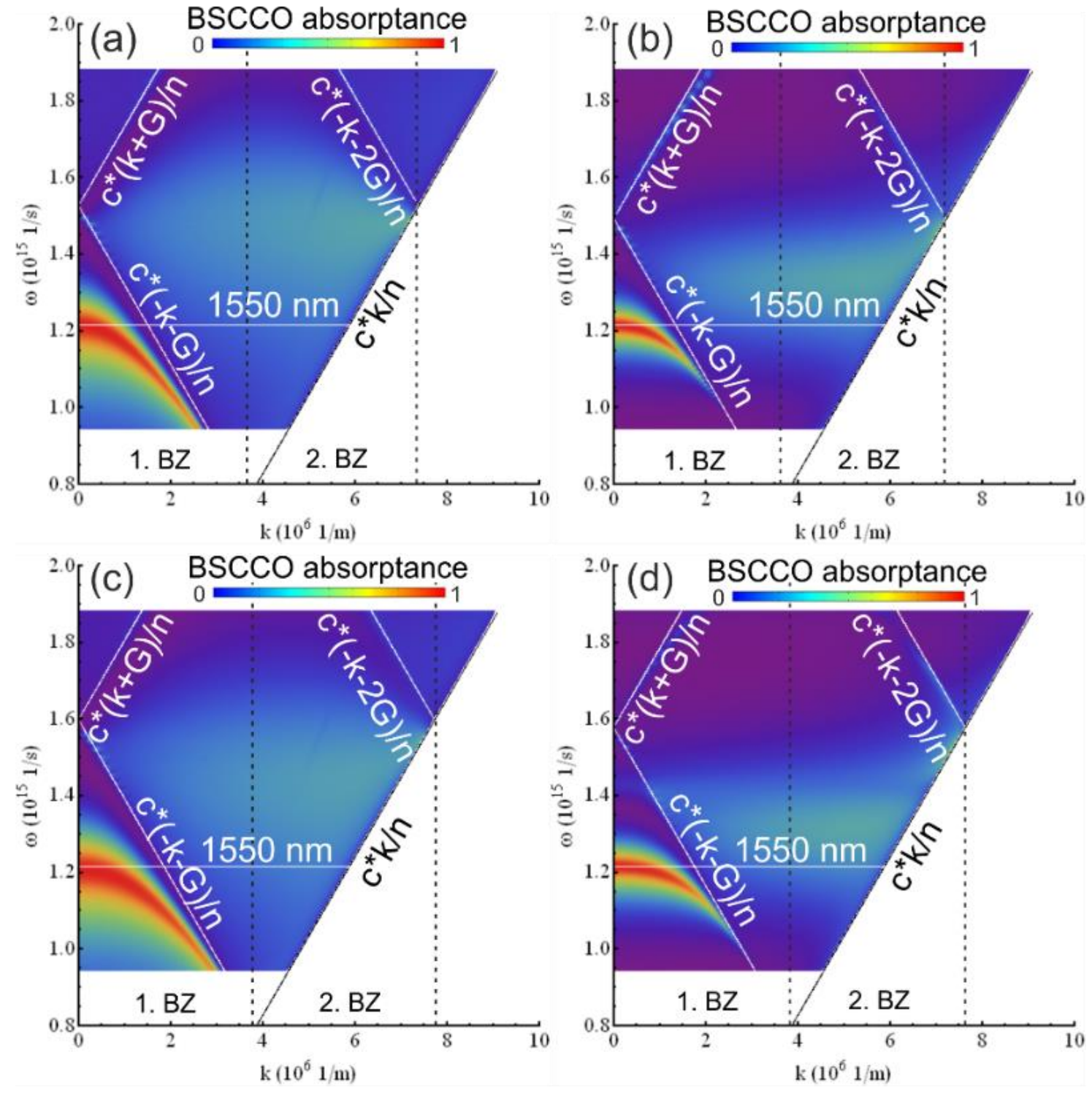

***Figure 4.*** *Dispersion diagram of BSCCO absorptance of (a) NCAI-SNSPD in λ/(4×n*$_{eff}$*) mode , (b) NCAI-SNSPD in 3×λ/(4×n*$_{eff}$*) mode, (c) NCTAI-SNSPD in λ/(4×n*$_{eff}$*) mode and (d) NCTAI-SNSPD in 3×λ/(4×n*$_{eff}$*) mode.*

Standard retrieval realized for perpendicular incidence reveals that the optimized nanocavity structure has an effective refractive index, which results in ~$\lambda/(4\times n_{eff})$ and ~$3*\lambda/(4\times n_{eff})$ at the first and third nanocavity resonance (Figure 3a, pink lines).

**3. Comparative study:** The absorptance spectrum and polar angle dependence of NCAI-SNSPD and NCTAI-SNSPD reveal similar resonant modes on the optimized integrated plasmonic gratings. To analyze the underlying physics, the dispersion diagrams of the optimized composition in p-to-S configuration are compared (Figure 4). The optimal NCAI-SNSPD composition exhibits well-defined low energy plasmonic mode below the first order grating coupling (Figure 4a and 4b). The branches are folded back into the first Brillouin zone (BZ) and result in a single Lorentzian resonance in the wavelength dependence at perpendicular incidence (Figure 2a). The absorptance remains large over a relatively wide polar angle interval (at incidence angles of 10° and 9°, the absorptance decreases to 1/e of its maximum value for the NCAI-SNSPD $\lambda/(4\times n_{eff})$, and $3\times\lambda/(4\times n_{eff})$ respectively) but sharply decreases below 0.01 around 13° tilting, which is in accordance with that the edge of this pass band is tuned to 1550 nm. Along the grating-coupled branches the absorptance of BSCCO is diminished, however at large tilting the absorptance is re-enhanced in a wide polar angle and wavelength interval across the second Brillouin zone. This corresponds to the plasmonic Brewster angle (PBA), which condition is met, when the grating is impedance matched with the substrate and results in anomalous energy squeezing into and tunneling through the nanocavity-array, hence results in suppressed reflectance and enhanced absorptance [30-31]. The maximum of PBA region is centered below 1550 nm. The PBA is calculated as (Table S4 and S5 in the Supplementary Materials) [30-31]:

$$PBA=acos(w_{cavity}\cdot\lambda/(n\cdot p\cdot\lambda_{plasmon}))$$

where $w_{cavity}$ is the width of cavity, $\lambda/n_{silica}$ is the incident wavelength in silica substrate, p is the period of the unit cell and $\lambda_{plasmon}$ is the wavelength of propagating plasmon polariton (SPP), which is approximated with the value calculated at the interface of semi-infinitemedia. It can be noted, that both the propagating plasmon's resonance and the PBA related peaks are significantly narrower in frequency in the optimized composition with $3\times\lambda/(4\times n_{eff})$ cavity length, which indicate high-Q modes of longer lifetime. This is also confirmed by eigenmode calculations which results in $\omega_{NCAI_1\times\lambda/4}$ =(1.24+0.09i)×10$^{15}$ 1/s and $\omega_{NCAI_3\times\lambda/4}$ = (1.22+0.05i)×10$^{15}$ 1/s (1522 nm and 1548 nm respectively) [32].

The optimal NCTAI-SNSPD compositions exhibit similar dispersion diagrams as their respective NCAI-SNSPD counterparts. A plasmonic mode below the first order grating coupling results in a pass band around 1550 nm and perpendicular incidence. In the second BZ a PBA phenomenon appears. In the case of $3\times\lambda/(4\times n_{eff})$ nanocavity length the resonance bandwidth is significantly smaller. The propagating plasmonic mode band is folded back into the first Brillouin zone, at similar frequency as in the case of NCAI-SNSPD ($\omega_{NCTAI_1\times\lambda/4}$ = (1.23+0.11i)×10$^{15}$ 1/s and $\omega_{NCTAI_3\times\lambda/4}$ =(1.22+0.06i)×10$^{15}$ 1/s), while the first order grating coupled SPP branch is observable in NCTAI-SNSPD as well.

Comparing the optimized integrated devices, the NCAI-SNSPD exhibits slightly larger BSCCO absorptance along with smaller reflectance than the NCTAI-SNSPD. The exact absorptance difference is 0.011 and 0.025 for the device operating in $\lambda/(4\times n_{eff})$ mode and $3\times\lambda/(4\times n_{eff})$ mode, respectively. While the NCAI-SNSPD possesses slightly higher BSCCO absorptance, it is achieved with a larger filling factor compared to the NCTAI-SNSPD. The latter, however, remains competitive in terms of electronic detector quantification because of the reduction of the charge crowding effect via shorter nanowire meander allows for a larger critical current [33].

supIn general, it can be concluded that the NCAI-SNSPD and NCTAI-SNSPD compositions exhibit very similar optical responses. This is because the geometry of the optimal systems is also very similar, since in optimized NCTAI-SNSPD configurations, the deflectors are so wide that only a narrow trench is created as a secondary period, i.e., the optimal NCTAI-SNSPD geometry approaches the NCAI-SNSPD geometry. The appearance of the secondary nanocavity inherently causes a difference in the optical response, so it subject of ongoing studies, whether a slightly smaller filling factor is optimal in NCTAI compositions.

The absorptance of BSCCO stirpes can be further enhanced by increasing their thickness to 30 nm (decreasing the encapsulating hBN thickness to 5 nm at the same time to keep the complete thickness constant) and re-optimizing the geometry. After that the main characteristics remain the same for all inspected compositions and configurations, namely a single plasmonic mode centered around 1550 nm appears and creates a pass band (Figure S6-S8 in the Supplementary Materials). With 30 nm BSCCO thickness the reflectance decreases below 0.01 while the absorptance reaches 0.867 and 0.798 in NCAI-SNSPD composition playing with $\lambda/(4\times n_{eff})$ and $3\times\lambda/(4\times n_{eff})$ modes, respectively, while 0.844 and 0.783 absorptance is attainable with NCTAI-SNSPD composition in $\lambda/(4\times n_{eff})$ and $3\times\lambda/(4\times n_{eff})$ resonant mode configurations, respectively.

**Conclusion:** In conclusion, the plasmonic grating integrated SNSPD based on BSCCO nanowire pattern is optimized to maximize absorptance at 1550 nm and perpendicular incidence. The nanocavity-array integrated SNSPD optical responses proves the capability to increase absorptance by more than an order of magnitude compared to a simple optimized optical nanocavity and an embedded meandered superconducting BSCCO pattern with a similar 0.1 filling factor. Both the $\lambda/(4\times n_{eff})$ and $3\times\lambda/(4\times n_{eff})$ scaled nanocavity-arrays are capable of maximizing absorptance, but the lower order cavity resonance in shorter nanocavity consistently results in larger BSCCO absorptance with smaller competitive gold absorptance and reflectance. While the NCAI-SNSPD possesses larger BSCCO absorptance, it is achieved with a slightly larger filling factor compared to NCTAI-SNSPD, which is competitive in the larger achievable critical current.

**Funding**

This work was supported by the National Research, Development and Innovation Office (NKFIH) of Hungary, projects: “Nanoplasmonic Laser Inertial Fusion Research Laboratory” (NKFIH-2022-2.1.1-NL-2022-00002) and “National Laboratory for Cooperative Technologies” (NKFIH-2022-2.1.1-NL-2022-00012) in the framework of the Hungarian National Laboratory program as well as Hu-rizont “Metamaterials for Optimized Quantum Information Processing” (METAOPTQIP, 2025-1.2.1-HU-RIZONT-2025-00107.

**Data availability**

All data generated or analysed during this study are included in this published article and its supplementary files.

# Supplementary Materials

# Plasmonic structure integrated superconducting nanowire single-photon detector constructed with BSCCO patterns

*András Szenes*[1,2] *, László Pothorcki*[1] *, Balázs Bánhelyi*[1,3]*, Mária Csete*[1,2]

[1] *Department of Optics and Quantum Electronics, University of Szeged, 6720, Dóm tér 9, Szeged, Hungary*
[2] *HUN-REN Wigner Research Center of Physics, Konkoly-Thege Miklós út 29-33, Budapest, Hungary*
[3] *Department of Computational Optimization, University of Szeged, 6720, Árpád tér 2, Szeged, Hungary*

## Material dispersion of BSCCO and hBN

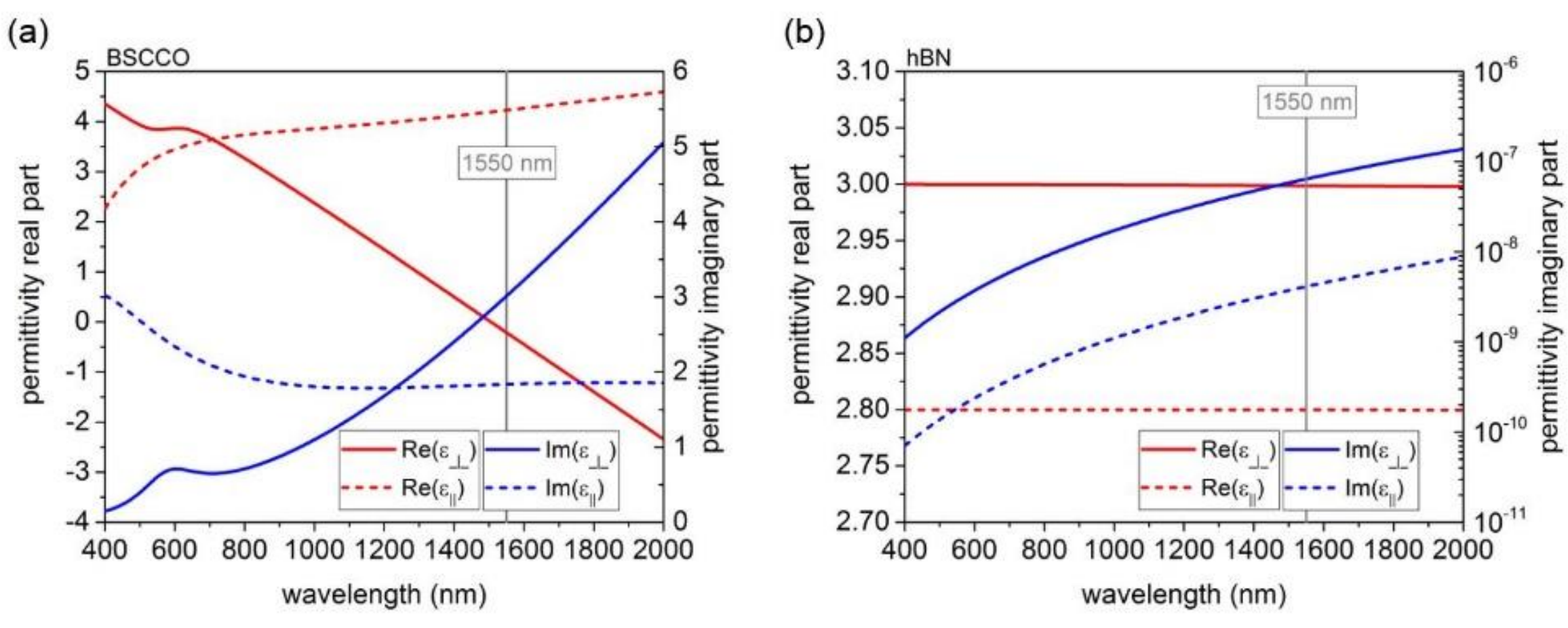

***Figure S1*** *Wavelength dependent dielectric permittivity of (a) BSCCO and (b) hBN used in the simulations.*

For BSCCO and hBN, their anisotropic optical properties were considered using a 3x3 tensor [S1, S2]. The elements of the complex diagonal tensor normal and parallel to the layer ($\varepsilon_\perp$ and $\varepsilon_\parallel$, respectively) are shown in Figure S1 in the wavelength range of 400-2000 nm. The imaginary part of the hBN is very small in this range, so its absorptance is close to zero, hence it is not plotted in the main text figures.

## Optical response of reference compositions

As a reference system for the presented NCAI-SNSPD and NCTAI-SNSDP devices, two simpler device compositions without integrated plasmonic structures were modeled.

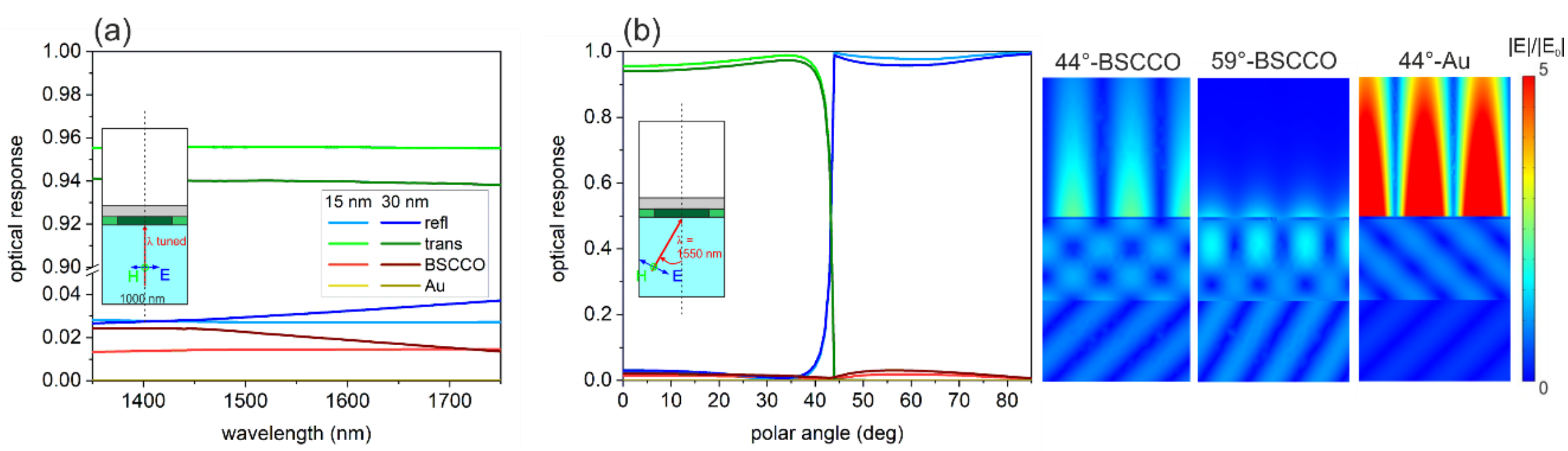

***Figure S2*** *Optical response of stand-alone hBN protected intact BSCCO wire and its leftover insulator (reBSCCO) complementary pattern on silica substrate (air is the bounding medium): (a) wavelength and (b) polar angle dependence. Insets: schematic of the geometry and configuration used for illumination with p-polarized light. Inset: evanescent a plasmon-like surface modes at the maximum in reflectance ($\varphi$=44°) and in absorptance ($\varphi$=59°) and propagating plasmon polariton mode on 35 nm thick gold slab for comparison.*

The absorptance of a bare continuous BSCCO layer as a function of wavelength and angle of incidence was investigated (Figure S2). For substrate side illumination, a low absorptance (less than 0.015,) was obtained, over the investigated wavelength range, though its value increases monotonically with wavelength. The reflectance is also very low, accordingly the dominant fraction of the incident power is transmitted. By varying the angle of incidence at 1550 nm the absorptance and reflectance decrease monotonically at small tilting and above the attenuated total reflection (ATR, at 44°) the absorptance shows a global maximum (at 59°), while the transmittance rapidly decreases to zero. At 44°/59°, an evanescent mode above/SPP-like propagating mode on the BSCCO layer is visible (Figure S2 insets). According to additional calculations, the near-field enhancement reached at the ATR of the BSCCO layer is significantly smaller compared to the near-field enhancement achievable at the SPP resonance of the gold thin film (Figure S2 insets). The commensurable near-field enhancement at the ATR and at the global absorptance maximum is the result of the BSCCO dielectric properties, which is differs from noble metals (Figure S1).

To enhance the absorptance, the BSCCO layer thickness has been increased to 30 nm. In that case not only the absorptance values, but the optical response tendencies also change. Namely, the absorptance is the largest at perpendicular incidence (0.021, Figure S2a) and monotonous decrease is observable. The polar angle dependence is similar with slightly increased absorptance throughout the complete inspected tilting range (Figure 2Sb).

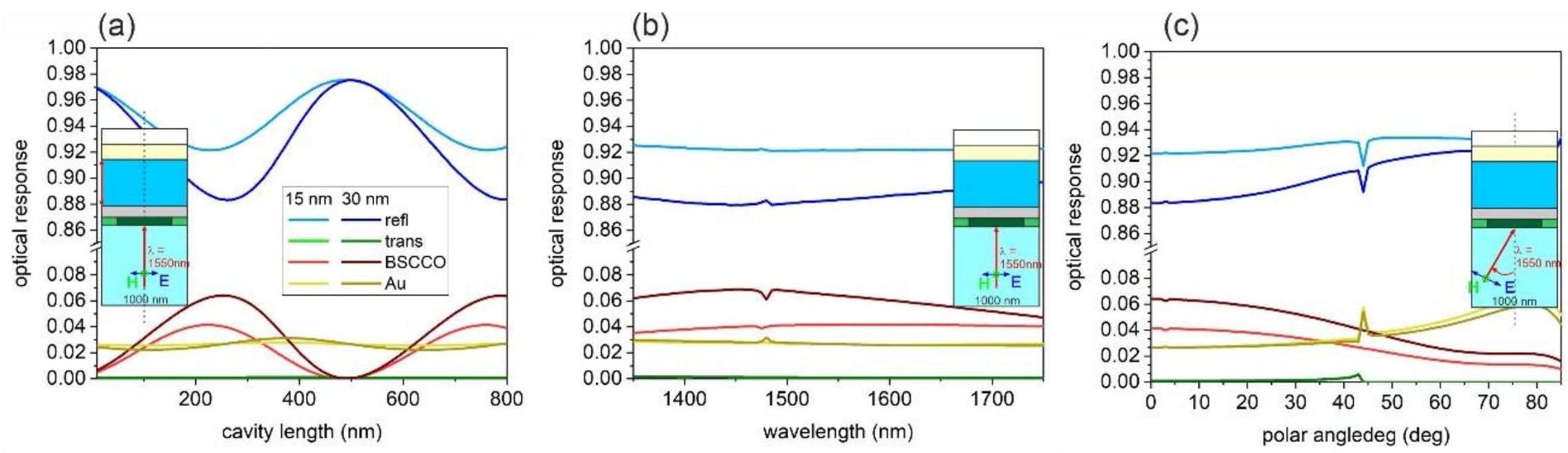

***Figure S3*** *Optical response of an optical cavity and gold reflector (60 nm) covering a hBN protected intact BSCCO wire and its leftover insulator (reBSCCO) complementary pattern: (a) cavity length tuning, (b) wavelength and (c) polar angle dependence. Insets: schematic of the geometry and illumination configuration used for illumination with p-polarized light in S-configuration (plane of incidence is perpendicular to the patterns).*

Next, a 60 nm thick gold reflector terminated the BSCCO layer covered with an HSQ spacer forming an optical cavity (OC). First, for perpendicular incidence and a fixed wavelength of 1550 nm, the optimal thickness of the dielectric spacer was determined, by maximizing the achievable absorptance (Figure S3a). It was found that thicknesses of 225 nm and 760 nm both provide higher absorptance of approximately 0.041, with high reminiscent reflectance, low gold absorptance and negligible transmittance. With fixed 225 nm cavity length the BSCCO absorptance monotonically increased due to the slightly /significantly increasing imaginary part of the parallel /normal component of the permittivity (Figures S1a and S3b) with decreasing complementary reflectance in the inspected wavelength interval. At 1550 nm wavelength, the impact of tilting was examined and it was shown that a monotonic decrease in absorptance is accompanied with a slightly and considerably increasing reflectance and gold absorptance, due to the gradually modifying reflection coefficient of the multilayer for the inspected p-polarized light.

By increasing the BSCCO thickness to 30 nm the optimal height of OC is increased to 255 and 785 nm. At perpendicular incidence, the absorptance is 0.064 and a global maximum is observable by increasing the wavelength with smaller reflectance and similar gold absorptance compared to the case of 15 nm BSCCO thickness. At 1550 nm, by changing the tilting, the BSCCO absorptance gradually decreases, whereas the gold absorptance increases, since the reflection coefficient of the multilayer for the inspected p-polarization was tuned. The absorptance is enhanced and the reflectance is suppressed compared to the 15 nm counterpart BSCCO film.

| | meander (15 nm) | | | meander (30 nm) | | | OC + meander (15 nm) | OC + meander (30 nm) |
|---|---|---|---|---|---|---|---|---|
| | 1550 nm, 0° | 1550 nm, 44° | 1550 nm, 59° | 1550 nm, 0° | 1550 nm, 43° | 1550 nm, 56° | 1550 nm, 0° | 1550 nm, 0° |
| BSCCO absorptance | 1.4 | 0.5 | 1.9 | 2.1 | 0.8 | 3.1 | 4.1 | 6.4 |
| Au absorptance | - | - | - | - | - | - | 2.7 | 2.7 |
| Reflectance | 2.7 | 99.5 | 98.1 | 3.1 | 99.2 | 95.9 | 92.1 | 88.3 |
| Transmittance | 95.6 | 0 | 0 | 94.0 | 0 | 0 | 0.1 | 0.1 |

***Table S1*** *Optical response of the inspected reference detector compositions. Geometrical parameters of the optical cavity + BSCCO layer are: 1000 nm BSCCO meander period and 225 nm and 255 nm optical cavity length for 15 nm and 30 nm BSCCO thickness, respectively.*

### Monte Carlo simulations of SNSPDs

As a first step of the optimization, Monte Carlo simulations were performed, i.e. the optical response of the device composition was calculated with random geometrical parameter values from a given range. In order to conclude, which geometrical parameter ranges could prove to be optimal, the absorptance read-out in intact BSCCO wire was plotted as a function of specific parameters.

The parameter bounds were as follows:

| NAME | INTERVAL (nm) | DESCRIPTION |
|---|---|---|
| $h_{HSQ}$ | 10 – 800 | height of dielectric cavity, excluding the 15+20 nm / 30 + 5 nm thick BSCCO +hBN bilayer |
| $w_{cavity}$ | 10 – 250 | width of nanocavity and intact superconducting BSCCO wire |
| period | 800 – 1200 | period of unit cell i.e. the intact BSCCO meandered pattern |

***Table S2*** *Name, interval and description of optimized parameters in case of NCAI-SNSPDs*

| NAME | INTERVAL (nm) | DESCRIPTION |
|---|---|---|
| $h_{HSQ}$ | 10 – 800 | height of dielectric cavity, excluding the 15+20 nm / 30 + 5 nm thick BSCCO +hBN bilayer |
| $w_{cavity}$ | 10 – 250 | width of cavity and intact superconducting BSCCO wire |
| period | 1000 – 1200 | period of unit cell i.e. the intact BSCCO meandered pattern |
| $w_{def}$ | 10 - (period-$w_{cavity}$)/2 | width of cavity walls (deflectors) |

***Table S3*** *Name, interval and description of optimized parameters in case of NCTAI-SNSPDs*

Initially the $h_{HSQ}$ height interval was set to 250-800 nm, which corresponds to a quarter wavelength to three-quarter wavelength nanocavity interval, i.e. 1550 nm / (4×$n_{silica}$) to 3×1550 nm / (4×$n_{silica}$). This is an approximation, since the effective optical parameters modify in presence of the absorbing layers. The period interval is centered around the wavelength of photonic and plasmonic modes that propagate in the medium, namely $\lambda_{photonic}$=1550 nm / $n_{silica}$ = 1068 nm and $\lambda_{plasmonic}$= 1055 nm, supposing semi-infinite bounding gold-silica layers. The nanocavity width covered a wide interval: from 10 to 250 nm.

These intervals were extended compared to the bounds in our previous works, considering the Monte Carlo results and effective wavelength calculations, which was necessary to allow reaching the global absorptance maximum.

Based on the nanocavity-dependent absorptance, indeed two modes can be excited in the NCAI-SNSPD compositions over the nanocavity length range under study (Figure S4). From the effective parameter retrieval analysis, it was found that these modes can be related to the modes of $\lambda/(4\times n_{eff})$ and $3\times\lambda/(4\times n_{eff})$ long cavities. The $\lambda/(4\times n_{eff})$ mode provides higher global maximum in the BSCCO absorptance. As the nanocavity width (and also the BSCCO fill-factor) increases, the absorptance increases, but changes only slightly above 110 nm width. As the period increases (and the BSCCO fill-factor decreases), the absorptance decreases slowly first and then decreases rapidly above 1000 nm. Above 1050 nm, local extrema appear, corresponding to the photonic mode.

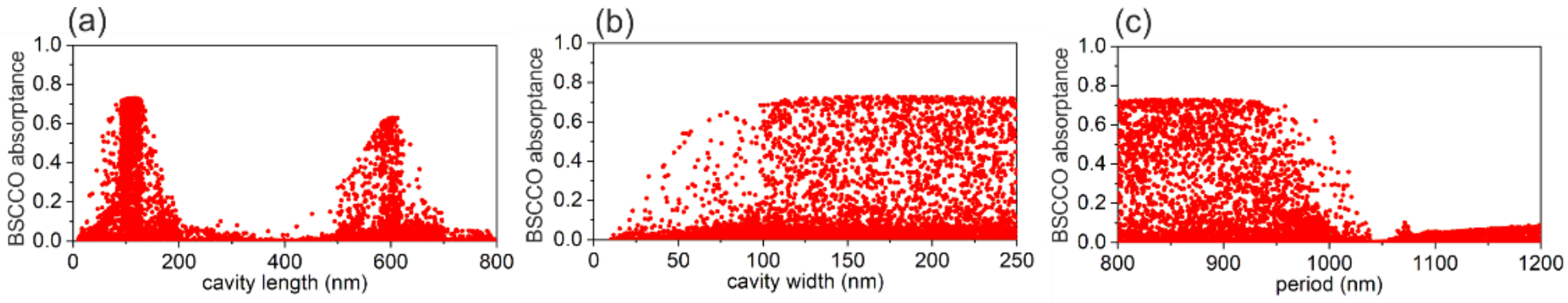

***Figure S4*** *Monte Carlo simulation of NCAI-SNSPD. Absorptance as a function of (a) nanocavity height, (b) nanocavity width and (c) pattern period.*

For NCTAI-SNSPD compositions (Figure S5), less well-defined maxima appear as a function of nanocavity length, due to the fact that a secondary nanocavity pattern is integrated onto the detector surface. However, it is similar in the two device compositions that the $\lambda/(4\times n_{eff})$ nanocavity modes provide the highest BSCCO absorptance. Interestingly, the period dependence of NCTAI-SNSPD is reversed compared to NCAI-SNSPD, since the absorptance takes on the global maximum at small periods. The nanocavity width shows similar dependence as in NCAI-SNSPD composition. For NCTAI-SNSPD compositions, the width of the nanocavity walls (deflectors) is also varying, which results in a global absorptance maximum around 400 nm, i.e. relatively larger amount of gold is optimal, which contradicts with the intuitive expectations. This may be related to the competing modes that does not play role and result in BSCCO absorptance enhancement.

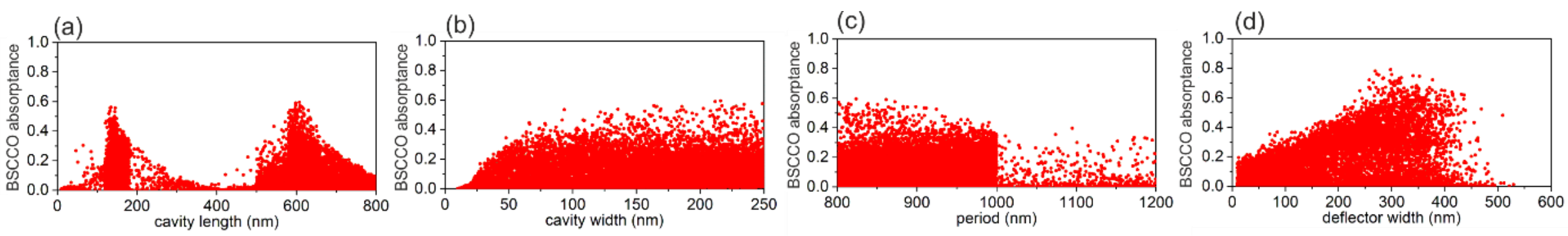

***Figure S5*** *Monte Carlo simulation of NCTAI-SNSPD. Absorptance as a function of (a) nanocavity height, (b) nanocavity width, (c) pattern period and (d) deflector width.*

## Optimized compositions

| optical response (1550 nm, 0°) | NCAI | | NCTAI | |
|---|---|---|---|---|
| | $\lambda/(4\cdot n_{eff})$ | $3\times\lambda/(4\cdot n_{eff})$ | $\lambda/(4\cdot n_{eff})$ | $3\times\lambda/(4\cdot n_{eff})$ |
| BSCCO absorptance | 0.726 | 0.626 | 0.715 | 0.601 |
| Au absorptance | 0.216 | 0.317 | 0.214 | 0.335 |
| reflectance | 0.039 | 0.041 | 0.059 | 0.054 |
| max. near-field | 5.4 (8.3) | 4.3 (7.3) | 5.4 (8.0) | 4.3 (6.9) |
| $n_{eff}$ (1550 nm) | 4.7 | 2.3 | 4.5 | 2.3 |
| $\lambda/(4\cdot n_{eff})$ (nm) | 82.4 | 165.6 | 86.4 | 167.7 |
| $3\times\lambda/(4\cdot n_{eff})$ (nm) | 247.3 | 496.8 | 259.1 | 503.0 |
| $\lambda/n_{eff}$ (nm) | 329.8 | 662.4 | 345.4 | 670.7 |
| $h_{HSQ}$ (nm) | 116.9 | 602.0 | 126.8 | 607.6 |
| total cavity height (nm) | 131.9 | 617.0 | 141.8 | 622.6 |
| period (nm) | 851.0 | 870.1 | 812.6 | 823.2 |
| cavity width (nm) | 173.2 | 241.4 | 161.2 | 216.0 |
| fill factor | 20.4% | 27.7% | 19.8% | 26.2% |
| PBA (°) | 78.1 | 73.7 | 78.4 | 74.6 |

***Table S4*** *Optical response of optimized SNSPD compositions at 1550 nm and perpendicular incidence and 15 nm thick BSCCO wire. The calculated effective refractive index and cavity height were also presented and the matched lengths are indicated with green background. PBA – calculated plasmonic Brewster angle.*

## SNSPD compositions optimized with 30 nm thick BSCCO pattern

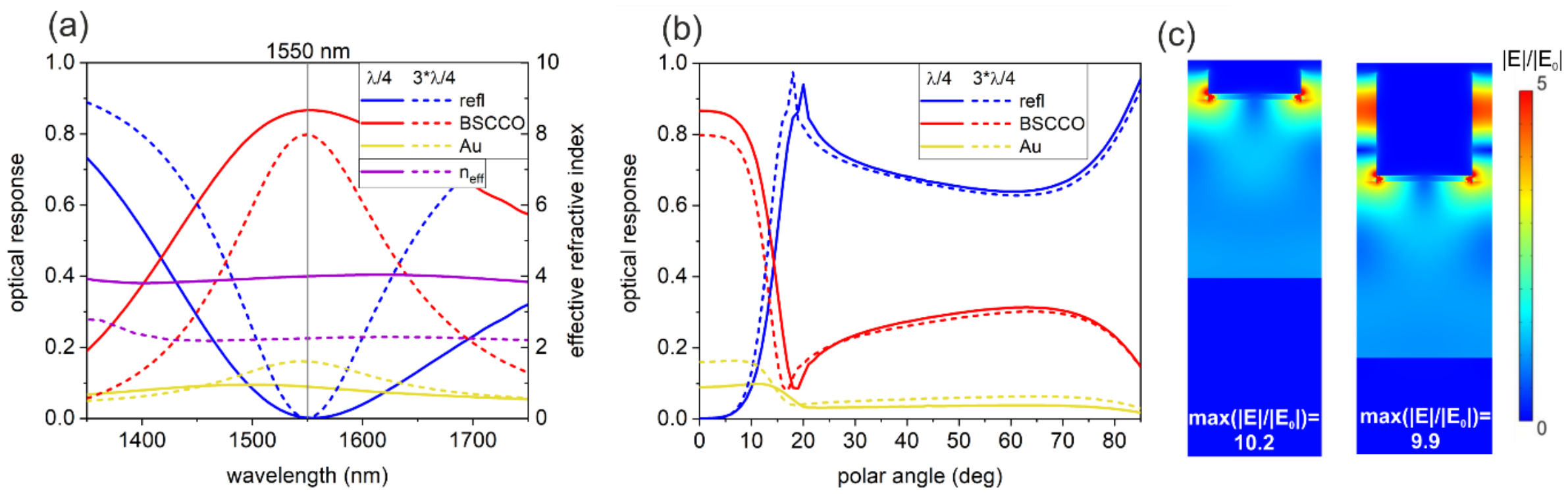

***Figure S6.*** *Optical response of optimized NCAI-SNSPD composition in p-to-S configuration. (a) optical response and effective refractive index spectrum and (b) polar angle dependence. Solid: ~$\lambda/(4\times n_{eff})$ mode dashed: ~$3\times\lambda/(4\times n_{eff})$ mode. (c) time-averaged near-field enhancement of NCAI-SNSPD optimized to resonate in the (left) $\lambda/(4\times n_{eff})$ mode and (right) $3\times\lambda/(4\times n_{eff})$ mode. Please note that the color scale was saturated at a value of 5 to enhance the visibility of fine structural details; the actual maximum near-field intensity exceeds this value.*

In the absorptance spectrum of NCAI-SNSPD compositions optimized with 30 nm thick BSCCO wire, a single resonance appears at the expected 1550 nm (Figure S6a). The optical response characteristic is similar to that of NCAI-SNSPDs optimized with a 15 nm thick BSCCO layer (Figure 2 in main text). It can be concluded that in the two configurations, the one with shorter cavities provides larger and broader absorptance maximum. By changing the polar angle (at 1550 nm), after a narrow plateau, the absorptance decreases rapidly in both configurations, then reaches a smaller local maximum at large angles (Figure S6b). The near-field and the effective refractive index also confirm that the nanocavity lengths are comparable to a quarter and three-quarters of the SPP modes wavelength propagating in the medium (Figure S6c).

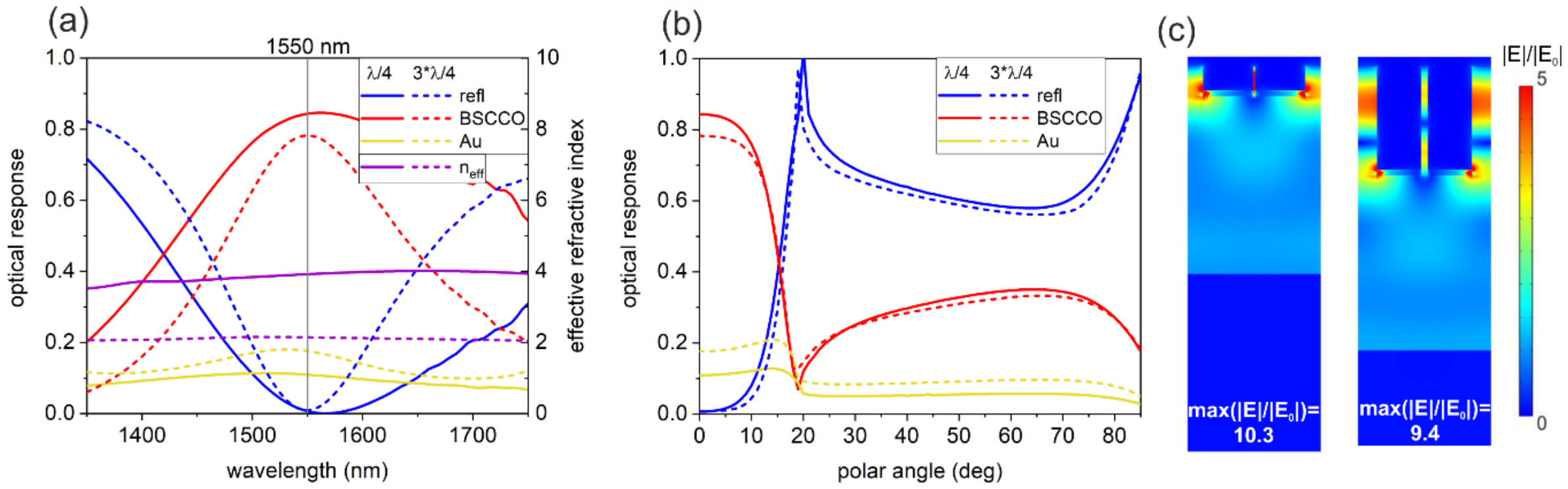

***Figure S7****. Optical response of optimized NCTAI-SNSPD composition in p-to-S configuration. (a) optical response and effective refractive index spectrum and (b) polar angle dependence. Solid: ~$\lambda/(4\times n_{eff})$ mode dashed: ~$3\times\lambda/(4\times n_{eff})$ mode. (d) time-averaged near-field enhancement of NCTAI-SNSPD optimized to resonate in the (left) $\lambda/(4\times n_{eff})$ mode and (right) $3\times\lambda/(4\times n_{eff})$ mode. Please note that the color scale was saturated at a value of 5 to enhance the visibility of fine structural details; the actual maximum near-field intensity exceeds this value.*

The NCTAI-SNSPD compositions optimized with 30 nm thick wires exhibit characteristics very similar to those of NCAI-SNSPDs (Figure S7). A resonant mode related single maximum appears at the examined wavelength of around 1550 nm (Figure S7a), the BSCCO absorptance is reaching a minimum at an incidence angle of around 20° and a local maximum at larger incidence angles (Figure S7b). Based on the near-field and effective refractive index, nanocavities comparable to a quarter and three-quarter wavelength proved to be optimal. Based on the geometry shown in the near-field figures, the NCTAI-SNSPD deflectors are so wide that the trench width approaches zero, i.e., it almost adopts NCAI-SNSPD geometry (Figure S7c). This explains the well-defined similarity between the optical responses of the two optimized compositions. The configuration with shorter nanocavities results in larger absorptance.

The two SNSPD compositions show very similar characteristics. Examining the BSCCO absorptance dispersion graph, it can be seen that a plasmonic mode propagates due to first-order lattice coupling, it band is folding back from the boundary of the first Brillouin zone (Figure S8). This plasmonic mode appears similarly in all compositions and configurations, proving that in all cases, tuning this coupled mode to 1550 nm results in the highest absorptance. The essential difference between the two configurations is that the resonance peak width is smaller with the longer nanocavity, which is consistent with the observation that the spectrum of higher-order modes is narrower. At the boundary of the first Brillouin zone and in the second zone, increased absorptance is observed over a wide range of frequencies and momenta. This is the result of the plasmonic Brewster angle, which is better localized above the ω-k parameter plane in the longer nanocavity configuration.

Greater absorptance can be achieved with quarter-wavelength nanocavity configurations, and among those, with NCAI-SNSPD compositions (Table S5). The highest absorptance is thus 0.867 in the quarter-wavelength NCAI-SNSPD configuration, which is only slightly lower than the 0.844 of the quarter-wavelength NCTAI-SNSPD. The absorptance in the three-quarter wavelength nanocavity length cases is 0.798 and 0.783 in the NCAI-SNSPD and NCTAI-SNSPD compositions, respectively. The advantage of NCTAI-SNSPD is that it achieves slightly lower absorptance with a smaller fill-factor.

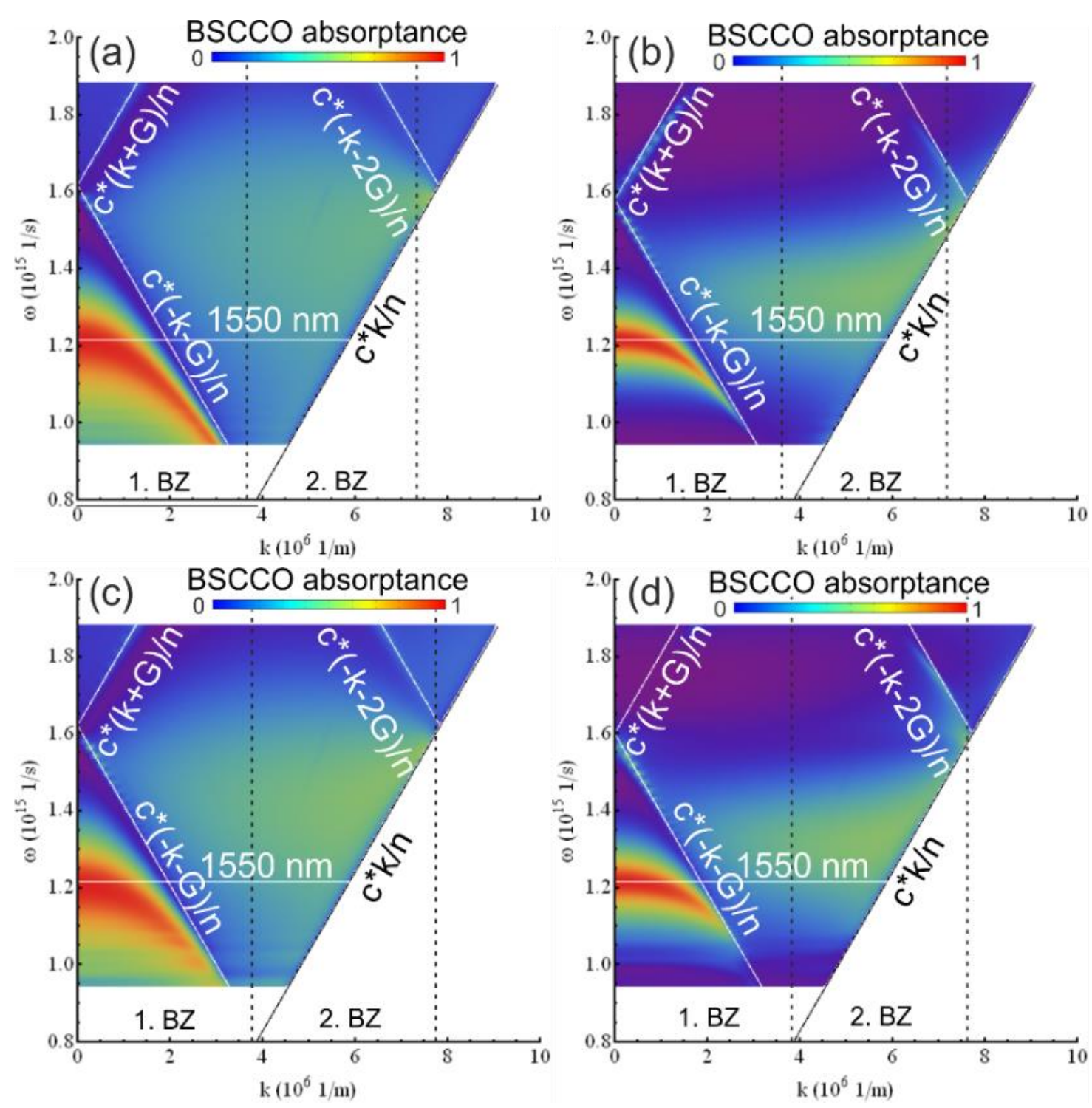

***Figure S8***. *Dispersion diagram of BSCCO absorptance of (a) NCAI-SNSPD in λ/(4×n<sub>eff</sub>) mode , (b) NCAI-SNSPD in 3×λ/(4×n<sub>eff</sub>) mode, (c) NCTAI-SNSPD in λ/(4×n<sub>eff</sub>) mode and (d) NCTAI-SNSPD in 3×λ/(4×n<sub>eff</sub>) cavity resonant mode.*

| optical response (1550 nm, 0°) | NCAI | | NCTAI | |
|---|---|---|---|---|
| | $\lambda/(4\cdot n_{eff})$ | $3\times\lambda/(4\cdot n_{eff})$ | $\lambda/(4\cdot n_{eff})$ | $3\times\lambda/(4\cdot n_{eff})$ |
| BSCCO absorptance | 0.867 | 0.798 | 0.844 | 0.783 |
| Au absorptance | 0.090 | 0.160 | 0.110 | 0.176 |
| reflectance | 0.002 | 0.001 | 0.007 | 0.009 |
| max. near-field | 3.4 (10.2) | 3.3 (9.9) | 3.8 (10.3) | 3.3 (9.4) |
| $n_{eff}$ (1550 nm) | 4 | 2.3 | 3.9 | 2.1 |
| $\lambda/(4\cdot n_{eff})$ (nm) | 96.9 | 171.4 | 98.6 | 181.0 |
| $3\times\lambda/(4\cdot n_{eff})$ (nm) | 290.6 | 514.2 | 295.8 | 543.1 |
| $\lambda/n_{eff}$ (nm) | 387.5 | 685.5 | 394.4 | 724.1 |
| $h_{HSQ}$ (nm) | 140.3 | 621.3 | 143.6 | 629.5 |
| total cavity height (nm) | 170.3 | 651.3 | 173.6 | 659.5 |
| period (nm) | 803.7 | 822.7 | 801.2 | 811.6 |
| cavity width (nm) | 246.0 | 241.8 | 185.4 | 231.5 |
| fill factor | 30.6% | 29.4% | 23.1% | 28.5% |
| PBA (°) | 72.0 | 72.7 | 76.5 | 70.4 |

***Table S5*** *Optical response of optimized SNSPD compositions at 1550 nm and perpendicular incidence and 30 nm thick BSCCO wire. The calculated effective refractive index and nanocavity height were also presented and the matched lengths are indicated with green background. PBA – calculated plasmonic Brewster angle.*

The advantage of the 30 nm thick BSCCO wire is that the absorptance can be significantly increased by more than +0.1 compared to the 15 nm thick wire.